\documentclass[journal=jacsat]{achemso}
\setkeys{acs}{keywords = true}
\usepackage[version=3]{mhchem}
\usepackage{graphicx}
\usepackage{xcolor}
\usepackage{setspace}
\graphicspath{ {./images/} }

\title{Surface microlenses for much more efficient photodegradation in water treatment}

\author{Qiuyun Lu}
\affiliation{Department of Chemical and Materials Engineering, University of Alberta, 9211 116 Street NW, Edmonton, Alberta, T6G 1H9, Canada}

\author{Qiwei Xu}
\affiliation{Department of Electrical and Computer Engineering, University of Alberta, 9211 116 Street NW, Edmonton, Alberta, T6G 1H9, Canada}

\author{Jia Meng}
\affiliation{Department of Chemical and Materials Engineering, University of Alberta, 9211 116 Street NW, Edmonton, Alberta, T6G 1H9, Canada}

\author{Zuo Tong How}
\affiliation{Department of Civil and Environmental Engineering, University of Alberta, 9211 116 Street NW, Edmonton, Alberta, T6G 1H9, Canada}

\author{Pamela Chelme-Ayala}
\affiliation{Department of Civil and Environmental Engineering, University of Alberta, 9211 116 Street NW, Edmonton, Alberta, T6G 1H9, Canada}

\author{Xihua Wang}
\affiliation{Department of Electrical and Computer Engineering, University of Alberta, 9211 116 Street NW, Edmonton, Alberta, T6G 1H9, Canada}
\email{xihua@ualberta.ca}

\author{Mohamed Gamal El-Din}
\affiliation{Department of Civil and Environmental Engineering, University of Alberta, 9211 116 Street NW, Edmonton, Alberta, T6G 1H9, Canada}
\email{mgamalel-din@ualberta.ca}

\author{Xuehua Zhang}
\affiliation{Department of Chemical and Materials Engineering, University of Alberta, 9211 116 Street NW, Edmonton, Alberta, T6G 1H9, Canada}
\email{xuehua.zhang@ualberta.ca}


\begin{document}
\begin{singlespace}

\maketitle
\clearpage

\section{Abstract}
\hspace{1.0em}
The global need for clean water requires sustainable technology for purifying contaminated water. Highly efficient solar-driven photodegradation is a sustainable strategy for wastewater treatment. In this work, we demonstrate that the photodegradation efficiency of micropollutants in water can be improved by $\sim$ 2-24 times by leveraging polymeric microlenses (MLs). These microlenses (MLs) are fabricated from the in-situ polymerization of surface nanodroplets. We found that photodegradation efficiency ($\eta$) in water correlates approximately linearly with the sum of the intensity from all focal points of MLs, although no difference in the photodegradation pathway is detected from the chemical analysis of the byproducts. With the same overall power over a given surface area, $\eta$ is doubled by using ordered arrays, compared to heterogeneous MLs on an unpatterned substrate. Higher $\eta$ from ML arrays may be attributed to a coupled effect from the focal points on the same plane that creates high local concentrations of active species to further speed up the rate of photodegradation. As a proof-of-concept for ML-enhanced water treatment, MLs were formed on the inner wall of glass bottles that were used as containers for water to be treated. Three representative micropollutants (norfloxacin, sulfadiazine, and sulfamethoxazole) in the bottles functionalized by MLs were photodegraded by 30\% to 170\% faster than in normal bottles. Our findings suggest that the ML-enhanced photodegradation may lead to a highly efficient solar water purification approach without a large solar collector size. Such an approach may be particularly suitable for portable transparent bottles in remote regions.

{\bf Keywords:}
surface microlenses, microlens array, solvent exchange, focus effect, photodegradation

\section{Introduction}
Earth receives abundant energy from the sunlight \cite{lindsey2009climate}. Solar-driven photodegradation is a sustainable and effective strategy for wastewater treatment.\cite{chong2010recent,chenab2020water} In some rural areas, especially remote regions, solar energy is one of the most popular resources to decontaminate water.\cite{mcguigan2012solar,moreno2021sodis} Compared to conventional wastewater treatment, photodegradation enhanced by catalysts \cite{dong2015overview} or sensitizers \cite{faust1999immobilized} can be highly efficient in removing pollutants that are hardly decomposed, such as antibiotics and personal care products.\cite{yang2020recent} However, currently low efficiency of light energy utilization and complexity associated with scaling up the size of solar energy collectors have hindered the applications of solar energy in fast conversion processes\cite{nakano2016remarkable, colmenares2017selective}. Innovative designs are needed to maximize the water decontamination efficiency of the incident solar energy per unit surface area for water purification.

Microlens (ML) is an optical element used in various fields for optoelectronic systems due to its strong directional control of the light field.\cite{cai2021microlenses, jin2014microlenses, lim2013micro} For instance, light extraction efficiencies of light emitting devices (e.g., organic light emitting diodes) are enhanced by MLs due to minimized total internal reflections. \cite{xia1998soft,jurgensen2021single, liu2019microlens} In solar harvesting systems, ML arrays are employed to improve power conversion efficiencies in solar cells \cite{fang2018antireflective, chen2013microlens}, and to focus light on the adsorbing material in solar evaporators for water desalination.\cite{dongare2019solar} However, the potential of MLs for wastewater treatment is restricted by not only the availability of a simple and low-cost approach for fabrication of MLs over a large surface area \cite{cai2021microlenses}, but also the understanding of effects of locally focused light on the photodegradation of organic contaminants in the aqueous matrix. 

Many bottom-up or top-down approaches have been developed for ML fabrication, such as hot embossing \cite{ong2002microlens,moore2016experimental} and laser writing technology \cite{gale1994fabrication,hua2020convex}. The techniques that offer precision in the morphology of MLs often require high energy consumption, or sophisticated setups, and hence are costly for large scale production of MLs for water treatment. \cite{daly2000microlens} Soft lithography is a low-cost and effective method, in which an elastomeric stamp or mold is used.\cite{kunnavakkam2003low,fang2018antireflective} However, the use of molds restricts the modification of ML shapes and positions, which makes it tedious to optimize MLs for desired optical properties.\cite{cai2021microlenses} Recently an interesting droplet-templating method has been reported where concave MLs are fabricated by covering curable polymers on templates of water microdroplets.\cite{mei2021facile} The curvatures of MLs are adjustable by varying the interfacial tension of water droplets at different cooling temperatures. Although water droplet templating is simple, flexible, and low cost, the size of the water droplets may be influenced by convection and heat transfer during droplet condensation, similar to the classic approach of `breath figure' templated porous structures \cite{bai2013breath}.

Conversion of surface nanodroplets to lenses is a newly established approach based on the nucleation and diffusive growth of nanodroplets in a solvent exchange process.\cite{zhang2012transient} Here the maximal height of the droplets is tunable from a few nanometers to 100 nanometers (namely nanodroplets). During the solvent exchange, the solution of droplet liquid in ethanol solution is displaced by water in a narrow channel. The oversaturation from the mixing between the solution and the poor solvent water leads to the nucleation and diffusive growth of the droplets on the wall surface. As most of the solvents are ethanol and water, the solvent exchange may be regarded as a green and environmentally-friendly approach with low energy cost for producing nanodroplets. To fabricate micro-/nanolenses, these surface nanodroplets which are composed of UV-curable monomers and an initiator are photopolymerized into lenses. The shape and positions of lenses are determined by the characteristics of surface nanodroplets, conveniently controlled by the flow conditions and solution compositions in the solvent exchange, and wettability and pre-patterns on the substrate surface.\cite{zhang2015formation, lu2016influence} Highly ordered ML arrays with tunable curvatures can be fabricated on a substrate with chemical patterns of hydrophobic circular microdomains.\cite{bao2015highly,bao2019control} MLs converted from surface nanodroplets can be adjusted with the volume of a ML as small as subfemtoliters and be uniform over an entire 12-inch wafer \cite{yu2016large}. 
 
The MLs fabricated from converting micron-sized droplets have been shown to exhibit a strong focusing effect in a total internal reflection mode. The local light intensity can be boosted up to $\sim$ 20 times, leading to a strong plasmonic effect of gold particles decorated on the lens surface.\cite{dyett2018extraordinary, bao2019control} Up to now, the study of chemical reactions enhanced by the MLs is limited to the local surface near MLs. It remains to be understood how focusing effects from heterogeneous and homogeneous MLs impact the photodegradation efficiency of organic compounds dissolved in an aqueous solution. 

In this work, we demonstrate that surface MLs fabricated from polymerized surface nanodroplets could significantly enhance the photodegradation efficiency of three real micropollutants and a model compound in aqueous solutions by up to 24 times. The optical properties of MLs were tuned by varying the size distribution, surface coverage, and patterning of precursor droplets to maximize the photodegradation efficiency. As demonstrated, MLs were fabricated on the inner wall of a glass bottle to boost the photodegradation efficiency of a model dye and three micropollutants in water. Up to 24 times higher efficiency was measured for the model dye, and up to 170\% for the micropollutants in water. Our work shows that surface MLs are promising to accelerate the solar-based photodegradation for fast wastewater treatment.

\section{Experimental section}
\subsection{Fabrication of the PMMA surface microlenses (MLs) on a planar substrate}
\begin{figure}
	\centering
	\includegraphics[scale=.55]{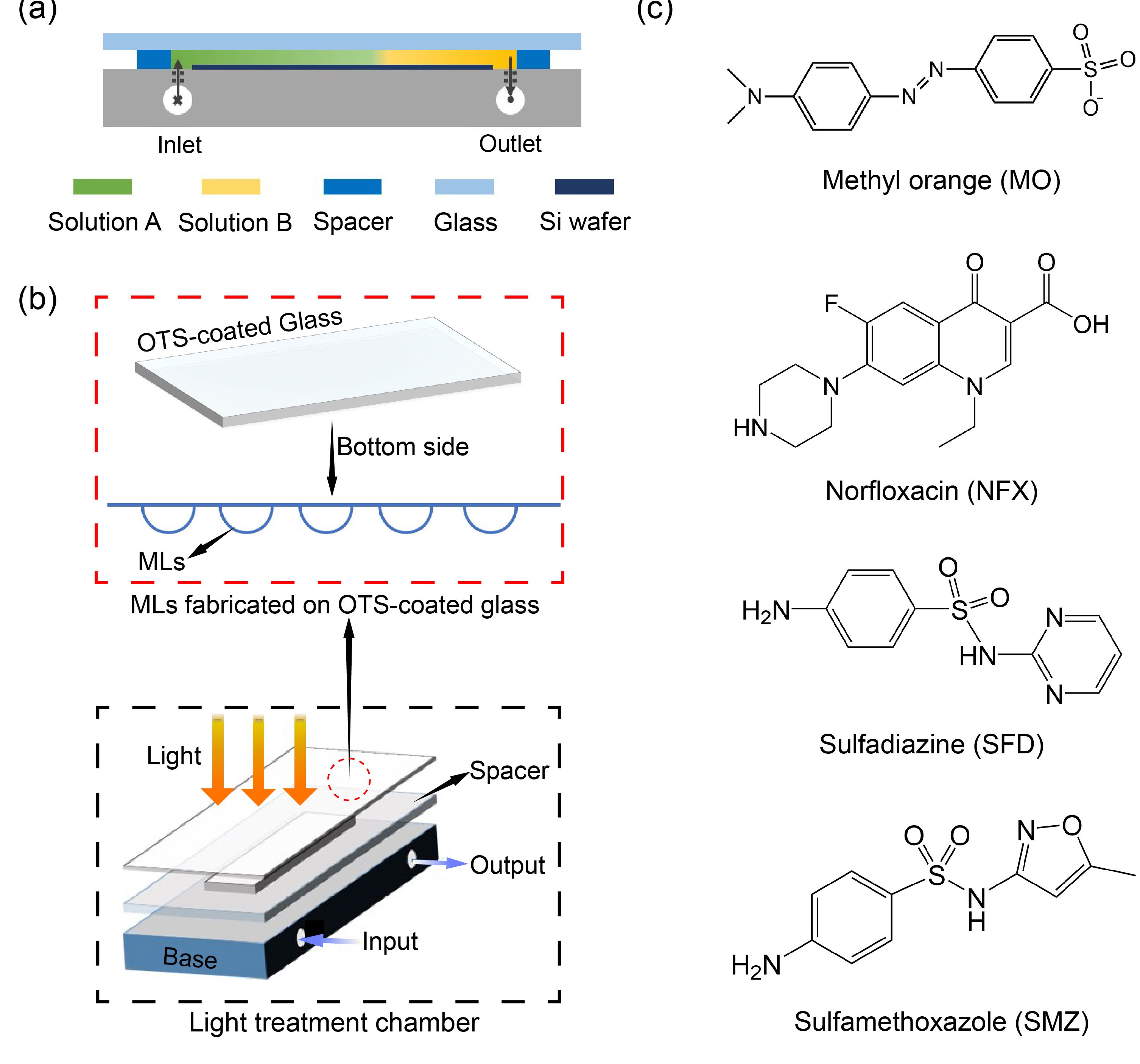}
	\caption{(a) The cross-sectional view of the flow chamber for solvent exchange (each component is labelled with different colors). (b) The schematic of the light treatment chamber for MO photodegradation with surface MLs.(c) Chemical structures of pollutants used in photodegradation experiments.}
	\label{Setup and chemicals}
\end{figure}

MLs of poly (methyl methacrylate) (PMMA) on planar substrates were prepared by following the procedure in the literature \cite{qian2019surface}. Briefly, the methyl methacrylate (MMA) droplets formed through a solvent exchange process, then the droplets were polymerized by UV light. In the solvent exchange process, the hydrophobic substrate, a glass slide ($25 mm \times 75 mm \times 1.0 mm$, Fisher Scientific) coated with octadecyltrichlorosilane (OTS) (98.9\%, Acros Organics, Fisher Scientific) was set on top of a customized chamber (height: 0.57 mm, width: 12.2 mm, length: 56.0 mm), as shown in Figure \ref{Setup and chemicals} (a).
The silicon wafer (thickness: 0.50 mm, University wafer) at the bottom was used to adjust the chamber height. The OTS-glass was firstly immersed into solution A, which was prepared using 3.5 mL MMA (98.5\%, Fisher Scientific) and 350 $\mu$L initiator 2-hydroxy-2-methylpropiophenone (96\%, Fisher) dispersed in water/ethanol solution (40 vol\% ethanol, 60 vol\% ultrapure water produced by with Milli Q system). Then, the MMA saturated ultrapure water solution (solution B) was pumped into the chamber with a fixed flow rate. The flow rate was controlled with a syringe pump (NE-1000, New Era Pump System). After solution A was replaced by solution B, the MMA droplets formed on the OTS glass substrate. The OTS glass with PMMA microlenses on the surface was rinsed with ultrapure water and ethanol for characterization and light treatment after removal from the chamber. 

Two types of MLs were prepared on the planar substrates with different wettability. The MLs fabricated on the homogeneous hydrophobic substrate were considered as random MLs due to the uncontrollable positions. The size of random MLs was adjusted by the flow rate of solvent exchange, and the flow rate was varied from 30 to 50, 70, and 90 mL/h. Another type of MLs was fabricated on a prepatterned substrate. The prepatterned substrate was fabricated with a reported protocol \cite{bao2015highly,choi2021effects}. Photoresist (AZ 1512) was spun coated on the OTS-coated glass substrate, and a photomask was subsequently attached to the substrate. The circular domains were protected by photoresist while the other area was etched with plasma during a typical lithography process. After removal of photoresist, the prepatterned substrate with hydrophilic background and hydrophobic circular domains with 5 $\mu$m diameter and 10 $\mu$m center-to-center distance was obtained. The ML array was obtained with a similar solvent exchange process and the following UV curing. In the solvent exchange process, the solution B (MMA saturated water) was added into solution A (4.5 vol\% MMA in 50 vol\% ethanol aqueous solution) with a flow rate of 8 mL/h. After the fabrication of PMMA ML array with one round of solvent exchange (termed as ML array\_1), the second round of solvent exchange with same solutions A and B was conducted to increase the height of MLs according to a method reported in the literature \cite{bao2019control}, and the flow rate was 5 mL/h when adding solution B. The ML array after two rounds of solvent exchange was represented as ML array\_2.

\subsection{Fabrication of the PMMA surface microlenses (MLs) on the inner wall of a bottle}
The MLs of PMMA were fabricated on the inner wall of a cylindrical glass vial with a volume of 30 mL (Fisherbrand Class A clear glass vial). The inner surface of the glass vial was first hydrophobized by an OTS layer. In the formation of MLs, solution A was prepared by adding 3, 4, or 5 mL MMA and initiator with the volume one-tenth of MMA in 60 mL 50 vol\% ethanol aqueous solution. Solution B was MMA saturated water containing 0.5 vol\% initiator. 

12 mL of solution A was put in an OTS-coated vial. 80 mL of solution B was added into the vial at the flow rate of 6 mL/min, while 62 mL of the mixture was taken out from the vial. After the solvent exchange process, the vial with droplets on the wall was put under the UV light for 30 min. The vial was rotated every 10 min to create uniform irradiation on the curved surface. After photopolymerization of the droplets, the vials with MLs were rinsed with water and ethanol subsequently, and labeled with MLs vial\_1, MLs vial\_2, and MLs vial\_3 correspond to the volume of precursors in solution A of 3 mL, 4 mL, and 5 mL.

\subsection{Characterization of surface MLs}
The morphology of MLs was examined using an optical microscope (Nikon H600l) equipped with a camera (Nikon DSFi3) and an atomic force microscope (AFM, Bruker, tap mode). The bottom area (S) and surface coverage of MLs were measured by analyzing more than 5 optical images from one sample (each image covering more than 1 mm$^2$) with ImageJ. With the bottom area of each microlens, the corresponding lateral radius R (R=$\sqrt{\displaystyle\frac{S}{\pi}}$) can be calculated. The cross-sectional profiles of MLs were extracted from AFM images. The contact angle of MLs, which referred to the angle at the contact line of MLs and substrate surface, was calculated according to the cross-sectional profiles. An integrated photovoltaic testing system (Sciencetech, PTS-2) was applied to measure the transmission (300-700 nm) of the OTS glass decorated with PMMA MLs. The transmission (unit: percent) was calculated as the ratio of transmitted light intensity through the samples to source intensity. The OTS glass without PMMA surface lenses was tested as the control group.

\subsection{Photodegradation of the model compound and micropollutants in water}
The performance of MLs and ML array on planar surface was evaluated with the photodegradation of model compound, methyl orange (MO, 85\%, Sigma-Aldrich), in a home-made light treatment chamber illustrated in Figure \ref{Setup and chemicals} (b). The bare OTS glass or MLs decorated OTS glass was put on top of the chamber, and the side with MLs was set towards the chamber. The light treatment chamber is 13.1 mm in width, 56.2 mm in length, and 3.05mm in height. The performance of MLs in light treatment was evaluated with different environmental factors, including pH value, initial concentration of MO, and dissolved oxygen level in solution. In this part, the MLs were prepared on homogeneous OTS substrates for convenience. The MO stock solution with different concentrations (3.5 mg/L, 5 mg/L, 7.5 mg/L, 20 mg/L, and 50 mg/L) was prepared with ultrapure water in advance and stored in a dark environment. The pH value of the MO stock solution was adjusted with sulfuric acid (98\%, Fisher) and measured with a pH meter (Accumet AE150, Fisher Scientific). To lower the dissolved oxygen level in the MO solution, the stock solution was degassed by an ultrasonic machine (degas mode) right before the light treatment. The dissolved oxygen level of the MO solution with and without the degassing step was detected with a dissolved oxygen probe (Model 50B, YSI Incorporated).

The MO solution was pumped into the light treatment chamber of which the inlet and outlet were then blocked. A simulated daylight LED with adjustable brightness (SOLIS-3C, Thorlabs) was set above the chamber with a fixed distance (23.5 cm) as the light source. The light intensity of irradiation at the upper surface of the chamber was measured with a miniature spectrometer (StellarNet Inc). After the light treatment, the MO solution was collected for subsequent characterization. The efficiency of photodegradation with and without MLs was measured after the light treatment of MO solution (initial concentration: 5 mg/L) for different time intervals, which can be calculated with the absorbance of the solution before and after light treatment. UV-vis spectrometer (Varian Cary 50) was utilized to obtain the absorbance value. The time of light treatment on MO solution was set as 15 min, 30 min, 45 min, 60 min, 240 min, and 480 min. The absorbance curves can provide information on the amount of MO because of Beer-Lambert law. According to Beer-Lambert law (equation \eqref{Beer Lambert law}), the absorbance at a certain wavelength is proportional to the concentration of solute: 
\begin{equation}
A = \log_{10}(\frac{I_0}{I})=\varepsilon CL
\label{Beer Lambert law}
\end{equation}
A is the absorbance of the sample, $I_0$ and I are respectively the light intensity before and after the light passing through the solution, $\varepsilon$ is the molar attenuation coefficient, C is the concentration of the analyte in the solution, and L is the length of the light path. Therefore, the ratio of the concentration of decomposed MO ($C_{i}-C_{f}$) to the initial concentration $C_{i}$, which is defined as the photodegradation efficiency ($\eta$) of MO, is calculated with the absorbance values by equation \eqref{Efficiency}.

\begin{equation}
\eta = \frac{C_{i}-C_{f}}{C_{i}}\times100\% =\frac{A_{i}-A_{f}}{A_{i}}\times100\% 
\label{Efficiency}
\end{equation}
$A_{i}$ is the absorbance at the representative peak \cite{zhang2006influence} of MO before light treatment and $A_{f}$ is the absorbance at the peak after the treatment.

Besides, an ultra-performance liquid chromatography-quadrupole time-of-flight mass spectrometry (Xevo $G_{2}$-S, Waters), operated in negative mode was used to analyze the by-products formed from the photodegradation at 30, 60, and 240 min. Chromatographic separation was achieved using ACQUITY UPLC BEH C18, 50×2.1 mm column, at 40 °C with an injection volume of 10 µL. The mobile phase consisted of water with 0.1\% formic acid (solvent A) and acetonitrile with 0.1\% formic acid (solvent B). The chemical structures of MO is shown in Figure \ref{Setup and chemicals} (c) and detected byproducts were listed in Figure S1 (a), which were consistent with the reported pathway of MO photodegradation \cite{he2011mechanism}. Electron spin resonance (ESR) spectra of the irradiated MO solution in Figure S1 (b) were obtained with a spectrometer (Elexsys E-500, Bruker), using 50 mM DMPO (5,5-Dimethyl-1-pyrroline N-oxide) as the spin trap.

The photodegradation of MO solution was conducted under the indoor solar light with the wavelength starting from 380 nm. The vials were fully filled with MO solution (5mg/L, pH=3.0) and were exposed to sunlight through a glass window for 15 days. The location of light treatment was Room 12-380 in Donadeo Innovation Centre for Engineering, Edmonton, Canada, and the experiments started from Nov 11 in 2021. The temperature was constant at 21 $^{\circ}$C.

Three types of micropollutants were photodegraded, including norfloxacin (NFX, Alta aesar), sulfadiazine (SFD, 99.0-101.0\%, Sigma Aldrich), and sulfamethoxazole (SMZ, analytical standard, Sigma Aldrich). The initial concentration of each pollutant in the solution was 5 mg/L. The vial was filled with the micropollutant solution and placed under the simulated solar light (1 sun, SS200AAA Solar Simulation Systems, Photo Emission Tech) for 4 hours. All samples after the light treatment were analyzed with UV-Vis spectrum to quantify the concentration of the degraded compound for the calculation of $\eta$. The enhancement of photodegradation efficiency was ($\eta_{MLs}$-$\eta_{bare}$)/$\eta_{bare}$. Here, $\eta_{MLs}$ is defined as the photodegradation efficiency with the MLs-decorated vial while $\eta_{bare}$ is the efficiency in the control.

\subsection{Optical simulations of surface MLs}
All optical simulations were performed in three-dimensional spaces using Zemax OpticStudio. MLs were placed in the same X-Y plane and illuminated by a plane wave source along the Z direction. The source intensity was set to be the same as in the experiment (21.64 W/cm2). In the MO solution, four light-flux monitors were placed in the X-Y plane at different Z positions to record light flux profiles at different depths. To demonstrate the focusing effect of MLs, X-Z plane monitors were placed along ML central axis to view cross-section intensity profiles.

\section{Results and discussion}
\subsection{Random MLs: morphology, size distribution, surface coverage, transparency and performance in MO photodegradation}
\begin{figure}
	\centering
	\includegraphics[scale=.43]{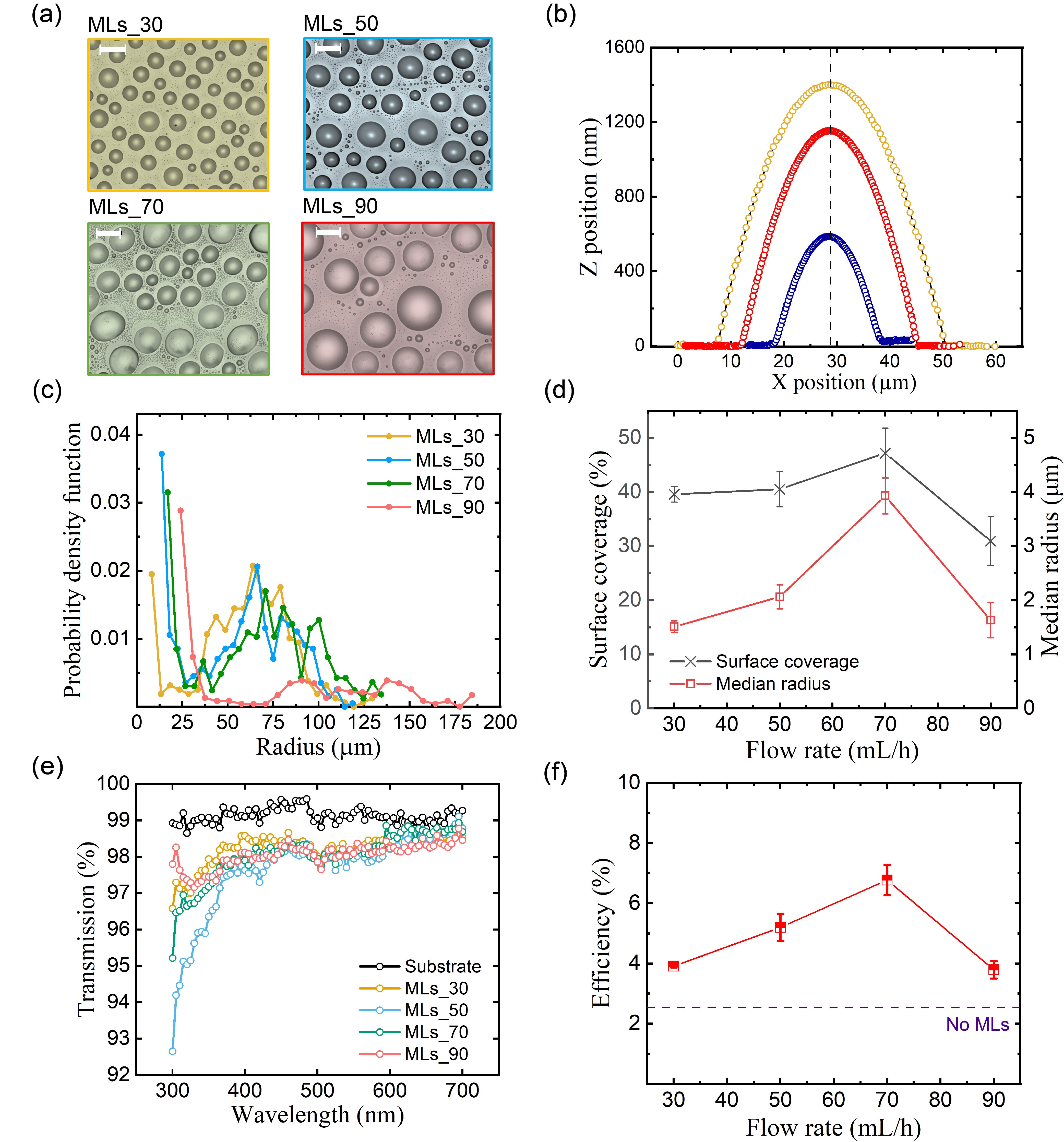}
	\caption{(a) Optical images of MLs fabricated with flow rates of 30 (MLs\_30), 50 (MLs\_50), 70 (MLs\_70) and 90 (MLs\_90) mL/h in solvent exchange process (scale bar 200 µm) (b) Cross-sectional profiles of MLs in homogeneous hydrophobic substrate by AFM. (c) Probability distribution function with MLs lateral radius. (d) The median lateral size and surface coverage of surface MLs fabricated with different flow rates. (e) The transmission of substrate (OTS glass) and the glass decorated by the surface MLs fabricated with different flow rates. (f) The photodegradation efficiency with MLs fabricated with different flow rates, and the dashed line represents the photodegradation efficiency without MLs. The MO solution used in photodegradation has the concentration of 5 mg/L at pH 3, and the irradiation time is 30 min.}
	\label{Properties of random MLs}
\end{figure}

MLs on the substrate are 3 $\mu$m to 200 $\mu$m in the lateral diameter and 1.5 $\mu$m in the maximal height (Figure \ref{Properties of random MLs} (a-b)).
The cross-sectional profiles of MLs in Figure \ref{Properties of random MLs} (b) extracted from atomic force microscopic images show that the aspect ratio of MLs is equivalent to a droplet with a contact angle around 7.5 (±0.2)°, similar to the morphology of microlenses fabricated by the same protocol in the literature \cite{lu2016influence}. The constant contact angle is a key morphological feature of MLs on the homogeneous surface, which is predetermined by the growth mode of the precursor droplets \cite{bao2016controlling}.

The properties of surface MLs can be tuned by altering the flow rate during solvent exchange.\cite{peng2014microwetting, zhang2015formation} The size distributions of surface MLs fabricated with flow rates of 30 ($MLs\_30$), 50 ($MLs\_50$), 70 ($MLs\_70$), and 90 ($MLs\_90$) mL/h are shown in Figure \ref{Properties of random MLs} (c). In each size distribution curve, the peaks of frequency are located in the range smaller than 25 µm and the range larger than 30 µm. With a higher flow rate, the peaks move to larger radius while the amount of MLs decreased. The phenomenon is consistent with the optical images in Figure \ref{Properties of random MLs} (a). In Figure \ref{Properties of random MLs} (d), the surface coverage continuously increases from 39.6\% to 47.2\% when the flow rate changes from 30 mL/h to 70 mL/h, and then decreases to 30.9\% at a flow rate of 90 mL/h. The trend of median lateral size of MLs with different flow rates is similar, reaching a maximum value of 3.93 µm at a flow rate of 70 mL/h.

The surface ML-decorated glass has high transparency in the visible light range as displayed in Figure \ref{Properties of random MLs} (e). According to the full spectrum of the LED lamp (Figure S2), the wavelength of the light source is between 400 nm and 800 nm. Within the wavelength range of the light source, the transmission of bare OTS-coated glass reaches 99\%, and the transmission of the MLs-decorated glass made with different flow rates is all over 97\%. Therefore, those surface MLs are adequate for the following light treatment, and the difference of transmission among the MLs prepared with different flow rates can be neglected. 

The photodegradation efficiency of MO after 30 min irradiation with random MLs prepared with different flow rates is plotted in Figure \ref{Properties of random MLs} (f). The photodegradation efficiency is obtained by inserting the absorbance peak value of MO solution before and after irradiation to equation \eqref{Efficiency}. The photodegradation efficiency with all random MLs is higher than the result obtained without using MLs. The efficiency increases with flow rate until the flow rate reaches 70 mL/h, and then drops at 90 mL/h. The maximum photodegradation efficiency is 6.8\% with the MLs made with a flow rate of 70 mL/h. The trend of photodegradation efficiency with flow rate is similar to that of surface coverage and median radius, indicating that the surface coverage and median size are possible factors that determine the performance of random MLs in the photodegradation enhancement. 

\subsection{ML arrays: 3D morphology and arrangement}
\begin{figure}
	\centering
	\includegraphics[scale=.43]{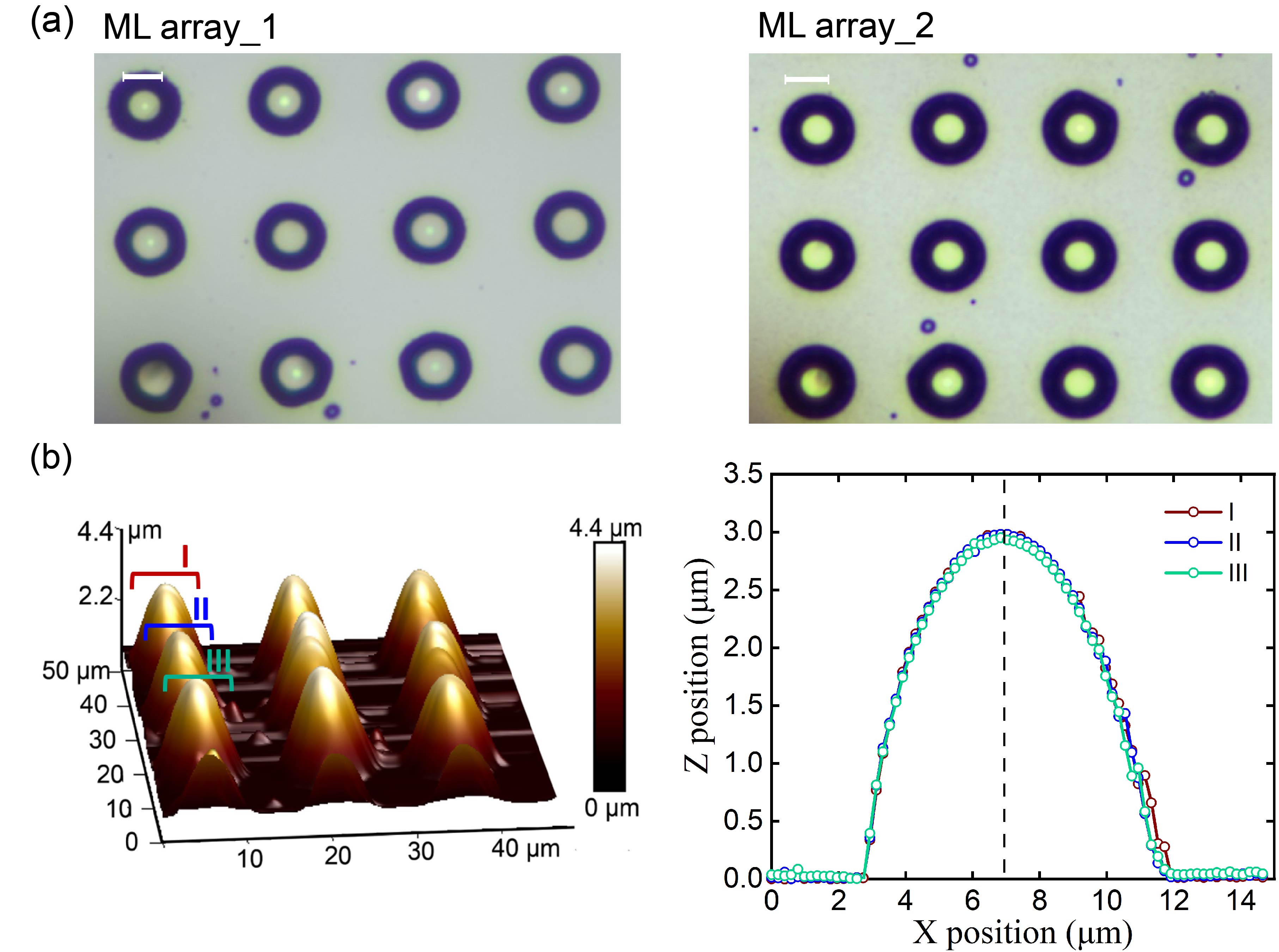}
	\caption{(a) Optical images of ML array fabricated with one round (left) and two rounds (right) of solvent exchange , and the contact angle of MLs in the array is 28 degree and 73 degree respectively (scale bar 5 µm). (g) 3D image of MLs after two rounds of solvent exchange by AFM and corresponding cross-sectional profiles.}
	\label{MLs array morphology}
\end{figure}

The MLs are arranged in a highly ordered array with uniform size on the prepatterned substrate, as shown in Figure \ref{MLs array morphology} (a). 
The MLs in the array prepared with one round (ML array\_1, Figure \ref{MLs array morphology} (a) left) and two rounds (ML array\_2, Figure \ref{MLs array morphology} (a) right) of solvent exchange have a similar lateral diameter around 9 µm, and the center-to-center distance is about 16 µm. The light spots in ML array\_2 are smaller but brighter than those in ML array\_1, and the difference in optical images indicates that the curvature of MLs further increased after the second round of solvent exchange \cite{lei2018formation}. The 3D image and the cross-sectional profiles of ML array\_2 in Figure 3 (c) show that the height of MLs in ML array\_2 reaches 3.0 (±0.1) µm and the contact angle is 73 (±0.5)° with two rounds of solvent exchange. The high curvature of MLs in the array is attributed to the confinement of circular hydrophobic domains.\cite{lei2018formation} The characterization results prove the feasibility of curvature adjustment of ML array with constant radius via multiple rounds of solvent exchange.

\subsection{Comparison of random MLs and ML arrays for photodegradation}
\begin{figure}
	\centering
	\includegraphics[scale=.43]{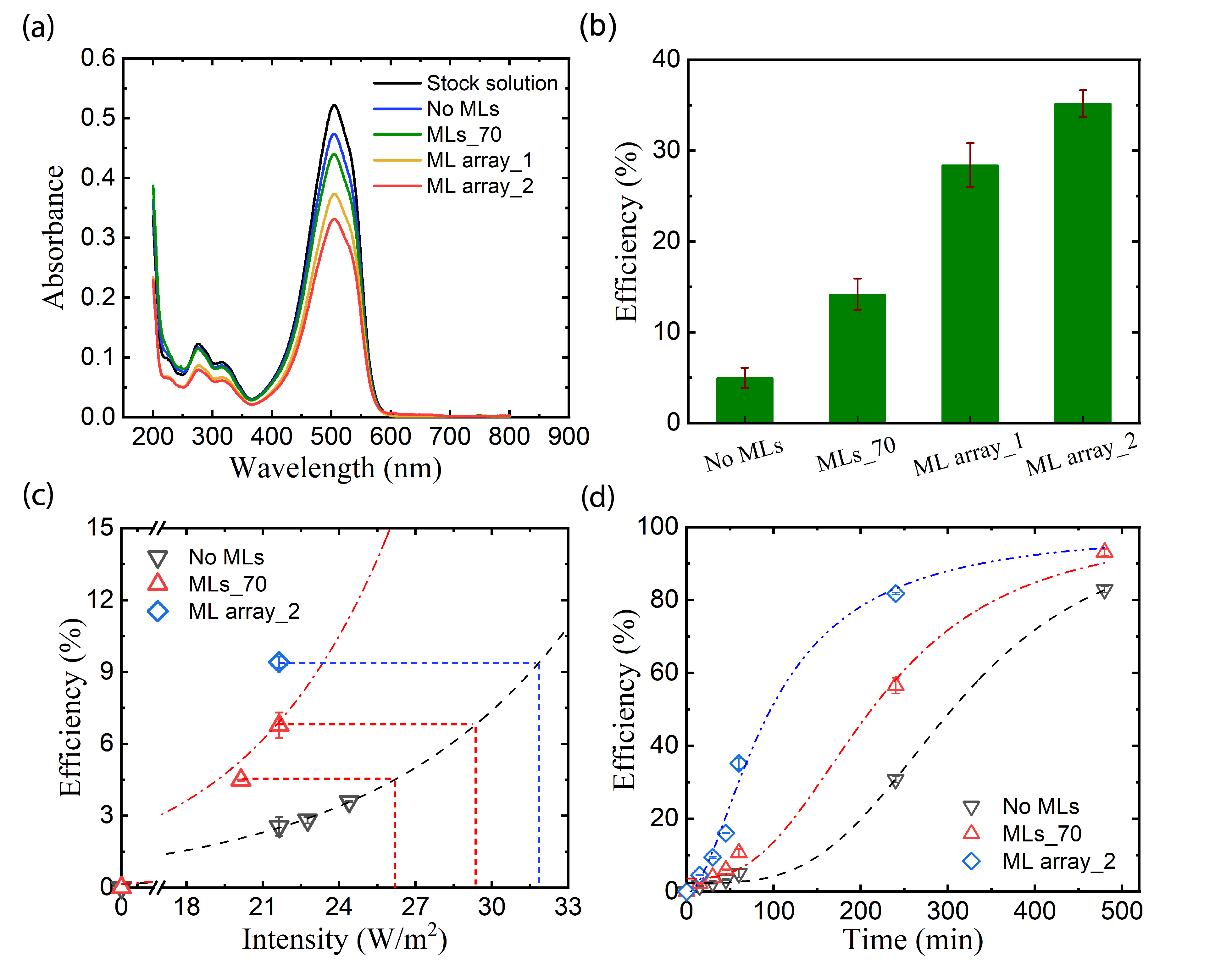}
	\caption{(a) Absorbance curves of methyl orange solution before and after 1 hour irradiation under different conditions. (b) Degradation efficiency of different types of surface MLs after light irradiation for 1 hour. (c) The photodegradation efficiency under the irradiation with different light intensity for 30 min. The red and blue dashed line is used to label the light intensity required without utilizing MLs to achieve the same photodegradation efficiency as MLs\_70 and ML array\_2 respectively. (d) The photodegradation efficiency with varied irradiation time, fitted with the logistic model. The concentration of MO solution is 5 mg/L, with pH=3.0 and degassed for 15 min, and the light intensity is 21.64 $W/m^2$ if not mentioned.}
	\label{Efficiency and intensity influence}
\end{figure}

The performance of ML arrays in the photodegradation of MO is compared with the random MLs that enhance photodegradation efficiency most ($MLs\_70$). As illustrated in Figure \ref{Efficiency and intensity influence} (a), the absorbance peak of MO solution (5 mg/L, pH=3.0) at 504 nm decreases after the light exposure for one hour, and the reduced peak value varies with the configurations of surface MLs.
By inserting the absorbance value at 504 nm to equation \eqref{Efficiency}, we can obtain the photodegradation efficiency with corresponding surface MLs (Figure \ref{Efficiency and intensity influence} (b)). During the light treatment for 1 hour, $MLs\_70$ enhances the degradation efficiency of MO by 186\% in comparison with the treatment without MLs. In comparison, the degradation efficiency of MO increases by 471\% and 607\% when applying ML array fabricated with one round (ML array\_1) and two rounds (ML array\_2) of solvent exchange, respectively. 

The performance of surface MLs was investigated under different light intensities. The full spectrum of the light source with four intensities that are applied in experiments is displayed in Figure S2. In Figure \ref{Efficiency and intensity influence} (c), the degradation efficiency during irradiation for 30 min is plotted against the light intensity. Under the irradiation with the same light intensity of 21.64 $W/m^2$, the photodegradation efficiency is 269\% higher with $MLs\_70$ and 165\% higher with ML array\_2 than that without using MLs. By fitting the curve of efficiency with the light source intensity, we can predict the photodegradation efficiency under higher light intensity. As shown in the red dashed line, the photodegradation efficiencies in the presence of $MLs\_70$ under the intensity of 20.14 and 21.64 $W/m^2$ are similar to those observed under 26.20 and 29.38 $W/m^2$ in absence of MLs. To achieve the same level of MO degradation with ML array\_2 at the intensity of 21.64 $W/m^2$, the light source should reach 31.82 $W/m^2$ without MLs based on the fitting results. By using ML array\_2, 47.0\% light energy is saved compared with the situation without MLs. The light energy required for the degradation of MO is reduced because the utilization of irradiation was more efficient through surface MLs, especially the ML array. The application of MLs significantly enhances the degradation efficiency under weaker irradiation, providing a potential solution for the light treatment of contaminated water under natural light sources. 

To further analyze the photodegradation process with surface MLs, the degradation efficiency calculated with equation \eqref{Efficiency} and absorbance values (obtained from Figure S3) is plotted against the treatment time in Figure \ref{Efficiency and intensity influence} (d). Throughout the irradiation time from 0 to 480 min, the degradation efficiency of MO with MLs\_70 is higher than that without MLs but lower than that with ML array\_2. By fitting the data in Figure \ref{Efficiency and intensity influence} (d) with the logistic model, it is found that the degradation efficiency showed a non-linear growth with treatment time under each condition. The photodegradation efficiency grows exponentially at the early stage of the reaction, and the growth slows down when the photodegradation efficiency is around 50\%. During the photodegradation in the first 100 min, it is obvious that the reaction rate with ML array\_2 is the fastest, followed by that with MLs\_70. The phenomenon confirms that the photodegradation can be accelerated by surface MLs, while the ML array is more efficient, especially in the first 100 min. To achieve the photodegradation efficiency of 80\%, it takes 457 min without MLs and 354 min with MLs\_70, and 214 min with ML array\_2. The usage of surface MLs and ML array effectively shortens the treatment time in MO photodegradation. 

\subsection{Influence of chemical composition in MO solution on photodegradation}
\begin{figure}
	\centering
	\includegraphics[scale=.37]{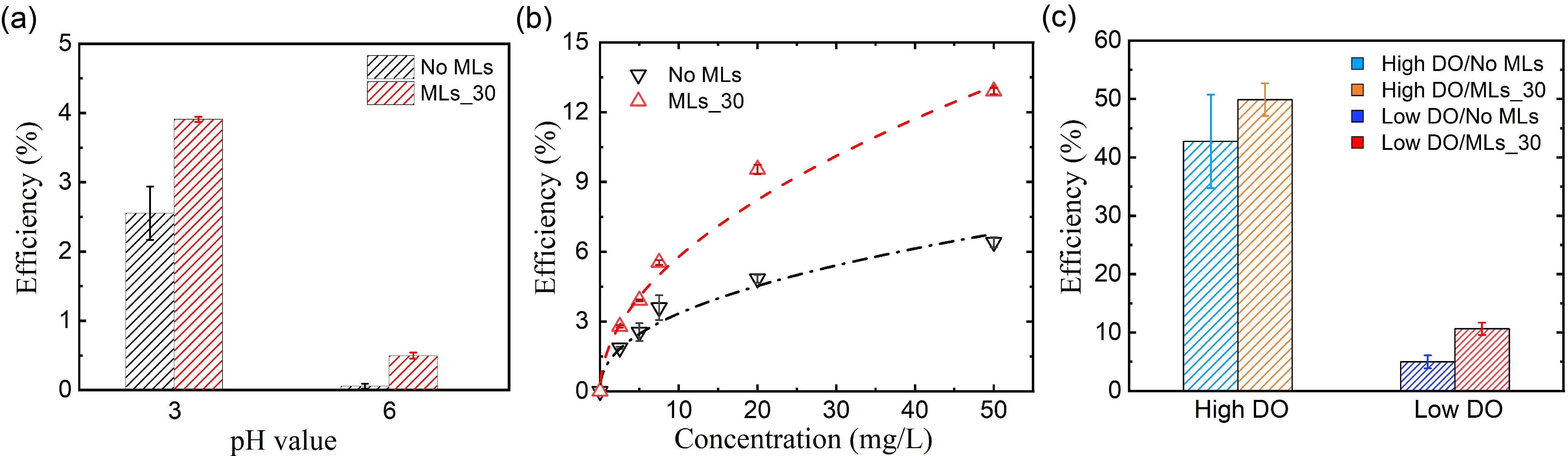}
	\caption{The photodegradation efficiency with (a) pH value and (b) concentration of MO solution without and with MLs (Fitting equation for the black curve: $y=1.22\cdot x^{0.44}$, the red curve: $y=1.79\cdot x^{0.51}$. (c) Photodegradation efficiency under different dissolved oxygen (DO) levels. The high DO level is 7.37($\pm 0.08$) mg/L, and the low DO level is 6.99($\pm 0.13$) mg/L.}
	\label{MO solution conditions}
\end{figure}

The pH value of the stock solution is one of the important factors affecting the photodegradation process. Figure \ref{MO solution conditions} (a) shows the efficiency of MO photodegradation both with MLs and without MLs under two pH values, 3 and 6. Without surface MLs, 2.6\% of MO is removed during the irradiation for 30 min at pH 3, while MO barely degrades at pH 6 under the same conditions.
The degradation efficiency is enhanced to 3.9\% at pH 3 and to 0.5\% at pH 6.0 when using the surface MLs. The possible reason for the higher degradation efficiency at lower pH is that the MO molecules dominantly exist in the protonated forms that are easier to photodegrade at pH 3.\cite{oakes1998kinetic} The surface MLs could accelerate the photodegradation of MO at both pH 3 and 6, and the effect is stronger at low pH value. Although the photodegradation efficiency is also improved by MLs at pH 6, the enhancement is limited because MO existed mainly as its inactive species at the pH value.

The effect of the initial concentration of MO solution on the performance of MLs was also investigated. Figure \ref{MO solution conditions} (b) displays the photodegradation efficiency of the MO solution with the initial concentration ranging from 2.5 to 50 mg/L after light treatment for 30 min. The trend of the efficiency with initial concentration is fitted with the power function shown in equation \eqref{Active species formation rate}, where $C_{0}$ is the initial MO concentration, a and b are the fitted variables.
\begin{equation}
\eta = a {C_{0}}^b\label{Active species formation rate}
\end{equation}
For the degradation process without MLs, the values of a and b are 1.48 and 0.39 respectively. For the fitting curve obtained with MLs, the value of a is 1.76 and that of b is 0.52. In other words, the degradation efficiency constantly increases with the MO initial concentration in the range between 2.5 and 50 mg/L. Furthermore, the photodegradation efficiency with MLs is consistently higher than that without MLs as the initial MO concentration varies. No matter the absence or presence of surface MLs, the enhancement of degradation efficiency slows down as the initial concentration increases, which is attributed to the limited amount of photons provided by the light source and the inhibition of light due to the high concentration of MO.\cite{reza2017parameters} However, the photons participate in the MO degradation more efficiently due to the light redistribution by using surface MLs. As a result, the photodegradation efficiency would further increase with the existence of MLs even though the initial concentration of MO reaches 50 mg/L.

The dissolved oxygen (DO) level in the MO solution is another factor that influenced the degradation efficiency. The high DO level is 7.37 ($\pm 0.08$) mg/L and is obtained from the MO solution without degassing step. The low DO level is 6.99($\pm 0.13$) mg/L and is from MO solution after degassing. With the absorbance curves before and after photodegradation (Figure S4) and equation \eqref{Efficiency}, the photodegradation efficiency without and with surface MLs under two DO levels is plotted in Figure \ref{MO solution conditions} (c). The photodegradation efficiency at the higher DO level is 38\% higher than that with the lower DO level. By applying MLs in the light treatment, the efficiency is enhanced on average by 7.1\% without degassing and by 5.7\% with degassing. The results confirm that DO could promote the photodegradation of MO.\cite{tasaki2009degradation,ren2016effects} However, the reproducibility of the experiments without the degassing step is worse compared with the tests under the lower DO level. We assume that the DO is not uniformly distributed in the sealed chamber because the DO is not in equilibrium without any mixing steps. This effect might be reduced by degassing the MO solution before filling the light treatment chamber with the solution. To guarantee the repeatability of results, the degassing step is done before the light treatment in all light treatment processes. 

\subsection{Optical simulations of surface MLs and ML arrays}
\begin{figure}
	\centering
	\includegraphics[scale=.46]{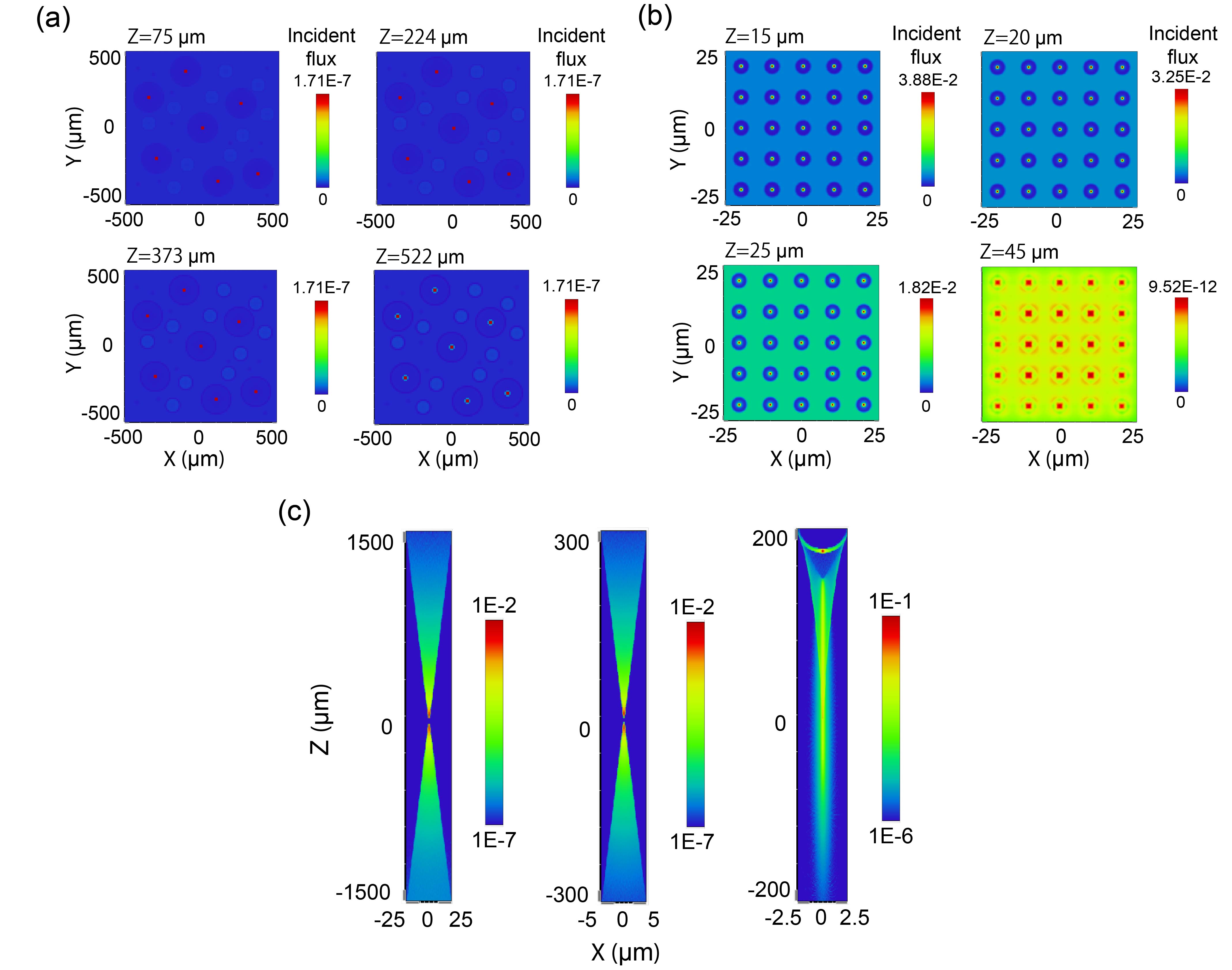}
	\caption{Top-view intensity profile of (a) random MLs at the depth of Z, and (b) ML array at the depth of Z. (c) Cross-sectional intensity profile of a single ML on homogeneous substrate with diameter of 50 µm and contact angle of 7.5° (left), on homogeneous substrate with diameter of 10 µm and contact angle of 7.5° (middle), and on a prepatterned substrate with diameter of 5 µm and contact angle of 73° (right).}
	\label{Optical simulation}
\end{figure}

To understand the effect of surface MLs in the whole light treatment chamber, the simulations of light paths through MLs on homogeneous hydrophobic substrates (random MLs, Figure \ref{Properties of random MLs} (a)) and ML array on a prepatterned substrate (ML array\_2, Figure \ref{Properties of random MLs} (f) right) are conducted. The top-view intensity profiles at the depths close to focal distances of MLs are demonstrated in Figure \ref{Optical simulation} (a-b). 
In the intensity profiles, the spots with higher light intensity are displayed with red color, indicating the focus effect of MLs. As shown in Figure \ref{Optical simulation} (a), only part of the focal points of random MLs are observed in the simulated region between the depth of 75 $\mu$m and 522 $\mu$m. The uneven size of random MLs causes that the focal points of MLs distributed at varying depths. It is also found that the light intensity at the displayed focal points of the random MLs almost remained similar from the depth of 75 $\mu$m to 522 $\mu$m.

Figure \ref{Optical simulation} (b) illustrates the top-view intensity profiles of ML array at the depth from 15 to 45 $\mu$m. The focal points of MLs in the array are located in the same plane. Different from the random MLs, rapid decay of light intensity is observed when the depth increased in the situation of the ML array. Even though the decay is much more distinguished below the focal points of the ML array, the highest light intensity achieved by the ML array is 5 orders of magnitude larger than that by random MLs. Therefore, 'hotter' spots are created at the focal points of MLs are fabricated on a prepatterned substrate. In addition, the focal points of MLs on homogeneous substrates are located at varying depths, but the focal points of ML array are densely located on a specific plane. The uniform focal distance of the ML array leads to a region with a high concentration of active species and accelerates the local photodegradation more efficiently than random MLs, which can be considered as a 'regional effect'.

At the same depth in the light treatment chamber, as displayed in Figure S5, the top-view intensity profiles change with surface MLs arrangement. At the depth of 224 $\mu$m, the highest light intensity in the situation with random MLs is much higher than that with ML array. The reason is that this depth is close to the focal distances of some large MLs on homogeneous substrates and is much larger than those of MLs on prepatterned substrates. Therefore, the region with strengthened light intensity is significantly influenced by the configuration of surface MLs. 

Based on the top-view intensity profiles obtained by optical simulations, we assume that the radius and curvature of MLs are the two main factors that influence the performance of MLs in photodegradation. As proof of the hypothesis, the cross-sectional intensity profiles of MLs with different radius and curvatures are simulated in Figure \ref{Optical simulation} (c). The MLs on homogeneous substrates have the same contact angle of 7.5°, and the focal distance increased from 0.3 mm to 1.5 mm as the diameter changed from 5 $\mu$m to 25 $\mu$m. Each ML on a prepatterned substrate has a larger contact angle of 73° and a diameter of 5 microns. The focal distance is only 17.5 $\mu$m but the light intensity is 10 times larger than that of MLs with similar lateral size on homogeneous substrates, which is attributed to the higher curvature of the MLs with array configuration. 

\section{Correlation between the intensity at focal points of MLs and photodegradation enhancement}
Enhanced photodegradation by using surface MLs may be rationalized by effects from focused light on the kinetics of photodegradation. After the light treatment for the time from 0 to $t_f$, the efficiency $\eta$ is determined by the initial concentration $C_{ini}$ and the final concentration $C_f$. 
\begin{equation}
\eta = \frac{C_{ini}-C_{f}}{C_{ini}}\times100\% 
\label{efficiency2}
\end{equation}

\begin{equation}
C_{ini}-C_{f} = \int_{0}^{t_{f}}{r(\lambda)}dt 
\label{cf}
\end{equation}

The treatment duration $t_f$ is the same when the efficiency $\eta$ is compared with and without MLs. According to the second law of photochemistry \cite{zepp1977rates,persico2018photochemistry}, the production rate of active species at a given time $t$ in an aqueous solution, $r(\lambda)$, is given by the equation as below. 

\begin{equation}
r(\lambda) = K \times \int_{\lambda_{min}}^{\lambda_{max}}{I_{\lambda}}{C_{\lambda}^{'}}d\lambda \label{r_lambda}
\end{equation}

\begin{equation}
C_{\lambda}^{'} = {C_{m} \varepsilon_{\lambda,m}}{\phi_{\lambda,m}} 
\label{constant}
\end{equation}

Here $K$ is a conversion constant, $C_{m}$ is the concentration of reactant $m$ that forms the rate-limiting species. $\lambda$ is the wavelength of the light source, ranging from the minimum ${\lambda_{min}}$ to the maximum ${\lambda_{max}}$. For the wavelength of $\lambda$, $I_{\lambda}$ is the intensity that drives the photodegradation, ${\varepsilon_{\lambda,m}}$ is the extinction coefficient, and ${\phi_{\lambda,m}}$, the quantum yield of $m$. For the model compound is the same in all our experiments, all the parameters in $C_{\lambda}^{'}$ except $I_{\lambda}$ can be considered to be the same.

\begin{figure}
	\centering
	\includegraphics[scale=.43]{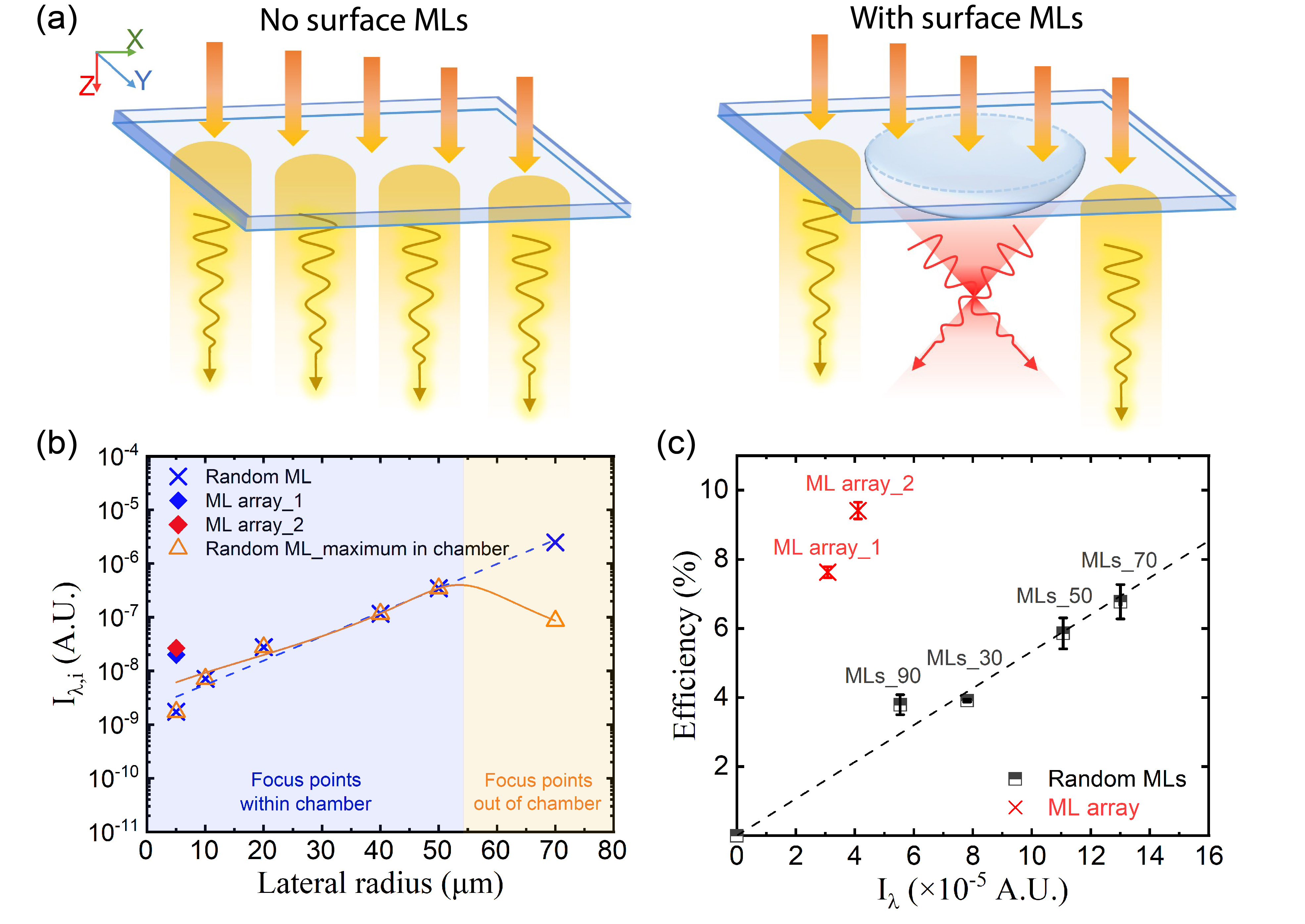}
	\caption{(a) Schematic of the light path without surface ML (left) and with surface ML (right) (b) Intensity ($I_{\lambda,i}$) at focal point of a single ML with different lateral radius. The intensity at the focal points of MLs is labelled with blue color, and the actual maximal intensity under random MLs is labelled with orange color. (c) The correlation between photodegradation efficiency and additive of intensity ($I_{\lambda}$) at focal points of MLs. For (b) and (c), the wavelength $\lambda$ is 504 nm in optical simulations.}
	\label{Influence of FP intensity}
\end{figure}

As sketched in Figure \ref{Influence of FP intensity} (a), in the case without using surface MLs, the light intensity $I_{\lambda}$ in equation \eqref{constant} is constant in a horizontal plane, and, for certain wavelength, decays with depth into the solution due to absorption by water. MLs alter the spatial distribution of $I_{\lambda}$ propagating in the solution. Beneath the surface areas without MLs, $I_{\lambda}$ is uniform and decays with the depth, same as the situation without MLs. Beneath the surface area with surface MLs, the light is concentrated on the focal points, resulting in a much higher $I_{\lambda}$ at the focal points. The above analysis shows that the spatially modified profile of $I_{\lambda}$ by MLs contributes to improved efficiency $\eta$ in the photodegradation. In water treatment by solar light, the size of the surface area that can receive the light is limited by the size of containers or the water reserve. Using MLs to redistribute the light in water may lead to more effective use of the surface area available. 

Below we will show that the enhancement from MLs is not simple additive, but exhibited coupled effects from multiple MLs. $I_{i}$ is the light intensity at the focal point of ML labelled as $i$. The total number of MLs over a certain area $A$ is $N$. 
\begin{equation}
I/A = (1/A_1) \times \sum_{i=1}^{N}{I_{\lambda,i}} + I_{\lambda}/A_2
\label{IA}
\end{equation}

Here $A_1$ is the substrate area occupied by MLs, and $A_2$ is the bare area. $I/A$ is the averaged intensity beneath the entire surface area $A$ (=$A_1$ + $A_2$). To simplify the comparison, we first neglect the second term ($I_{\lambda}/A_2$) on the right.

\begin{equation}
I_{\lambda} \approx \sum_{i=1}^{N}{I_{\lambda,i}}
\label{I_lamda}
\end{equation}

The energy intensity per unit area $I_{\lambda,i}$ created by individual ML(i) is obtained from optical simulation. The plots in Figure \ref{Influence of FP intensity} (b) show that $I_{\lambda,i}$ increases with the base radius of the MLs on homogeneous surfaces. For a given base radius, $I_{\lambda,i}$ is much higher from MLs in the array, thanks to their higher curvature. We note that $I_{\lambda,i}$ decreases for MLs with a radius larger than 54 $\mu$m, because the focal point of such large MLs is further than 3 mm, the depth of our reactor. The actual $I_{\lambda,i}$ contributing to photodegradation deviates from the simulated value for large MLs. A fitting line (the orange curve in Figure \ref{Influence of FP intensity} (b)) is used to predict the irradiation power at the focal point or hot spot of a ML on homogeneous hydrophobic substrates.

Considering both $I_{\lambda,i}$ from the simulation results and the morphology and number of MLs extracted from optical images, we calculate $I_{\lambda}$ of homogeneous substrates functionalized with MLs. Figure \ref{Influence of FP intensity} (c) shows that the efficiency $\eta$ increases with $I_{\lambda}$, and reaches the highest by MLs on homogeneous substrate fabricated with the flow rate of 70 mL/h. The reduced performance from MLs prepared at 90 mL/h is attributed to the energy loss due to the focal distance larger than the depth of the reactor. The overall efficiency $\eta$ is approximately linear with the increase of the sum of $I_i$ from all MLs obtained from homogeneous substrates. 

Interestingly, the efficiency $\eta$ from the ML array is 100 \% higher than random MLs with the same $I$, and is 39\% higher than the best results from random MLs on homogeneous substrates. Such high efficiency is much beyond the cumulative effect of $I_i$, as observed for homogeneous substrates. We propose that the photodegradation with MLs in an array could be attributed to the ordered arrangement of focal points on the same plane. Such ordered focal points may lead to a narrow distribution of spatial distribution of active species in water. The focal points on the same plane are also synchronized in certain zones of the aqueous solution. Such focusing effect increases the local concentration of active intermediate products and speeds up the photodegradation rate. In future work, even higher $\eta$ may be achieved as $I_i$ of microlens array is optimized. For instance, making closely arranged patterns increases the surface coverage, and the size of ML in the array can be tuned by the size of each domain.

\section{Proof-of-concept: MLs-enhanced photodegradation for water treatment}
\begin{figure}
	\centering
	\includegraphics[scale=.52]{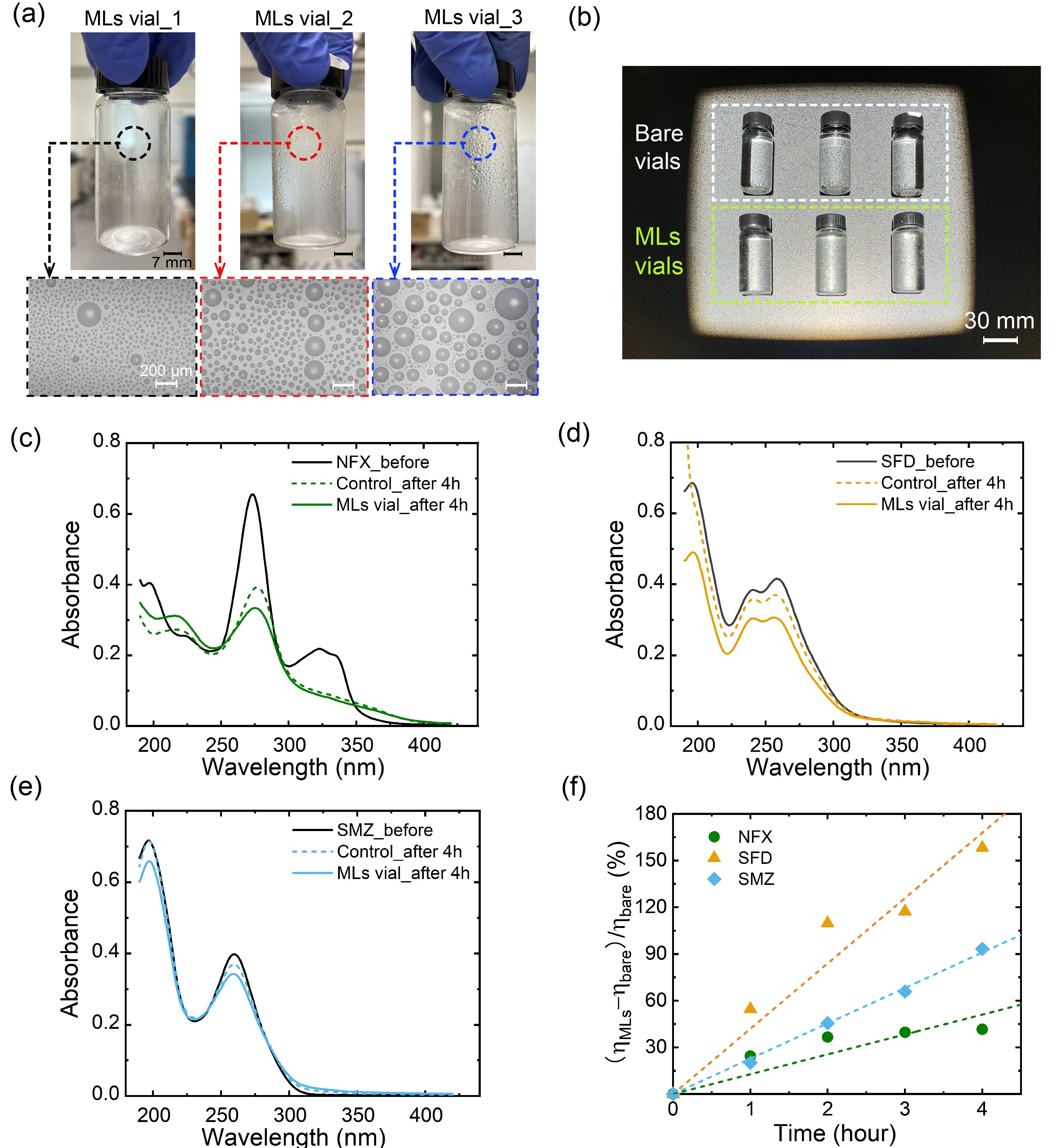}
	\caption{(a) Photos of the glass bottles functionalized by MLs. The bottles are labelled as MLs vial\_1, 2, and 3. The optical microscope images of the zoom-in areas show the MLs on the inner wall of the bottles. (b) Experimental set-up of simulated solar light treatment of micropollutant aqueous solutions. In the bare vials, from left to right: NFX, SFD, and SMZ. In ML-coated vial MLs vial\_3s, from left to right: NFX, SFD, and SMZ. Absorbance spectrum of (c) NFX, (d) SFD (e) SMZ solution (5 mg/L) before and after the exposure to simulated solar light for 4 hours. Dashed lines: solutions in bare vials; Solid lines: solutions in functionalized vials. (f) Enhancement of photodegradation efficiency of micropollutants as function of treatment time. $\eta_{MLs}$ is defined as the photodegradation efficiency with the MLs-decorated vial, while $\eta_{bare}$ is the efficiency in the control.}
	\label{MPs simulated solar treatment}
\end{figure}

The potential application of MLs in the photodegradation of harmful compounds in water is demonstrated by using functionalized bottles as shown in Figure \ref{MPs simulated solar treatment} (a). The photos show these bottles with the inner wall coated with MLs.

\begin{figure}
	\centering
	\includegraphics[scale=.55]{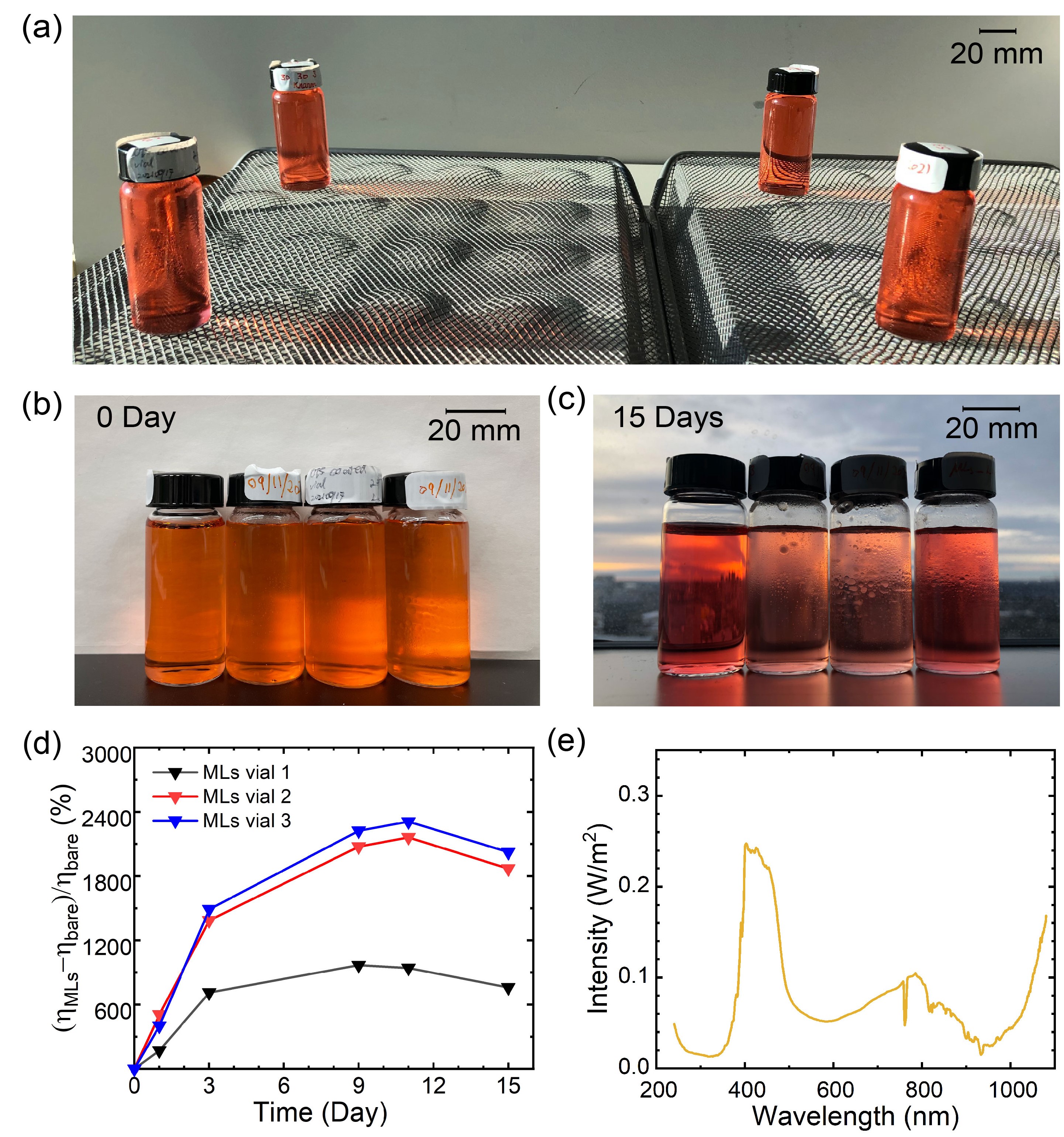}
	\caption{(a) Layout of four bottles receiving the indoor sunlight. Photos of the bottles filled with the dye solution (b) Before and (c) after receiving sunlight for 15 days. From left to right: bare vial, MLs vial\_3, MLs vial\_2, MLs vial\_1). (d) Enhancement of photodegradation efficiency of MO with the treatment time.(e) Spectrum of indoor solar light at the location of light treatment.}
	\label{office solar treatment}
\end{figure}

Three micropollutant solutions in the bottles were treated by the simulated solar light with the set-up in Figure \ref{MPs simulated solar treatment} (b). Each type of micropollutant solution is separately put in a bare vial and the MLs vial\_3. The difference in the micropollutant concentrations at a given time was detectable in UV-Vis spectra as shown in Figure \ref{MPs simulated solar treatment} (c-e)). The enhancement of photodegradation efficiency is plotted as a function of irradiation time in \ref{MPs simulated solar treatment} (f). After 1 hour, $\eta_{MLs}$ was already clearly higher than $\eta_{bare}$ by approximately 20\%. After the irradiation for 4 hours, $\eta$ of NFX was improved by 1.2 times by MLs, the lowest among the three micropollutants. $\eta$ of SFD and SMZ increased by 2.6 and 1.9 times, respectively. The results verify that surface MLs could increase the photodegradation efficiency of micropollutants in water.

The enhancement of photodegradation by MLs is also demonstrated with the light source of indoor sunlight. The dye aqueous solutions in the bottles were exposed to sunlight through a glass window (Figure \ref{office solar treatment} (a)). The short wavelength in natural solar light was cut off by the thick glass panels. As illustrated in Figure \ref{office solar treatment} (b) and (c), the color of the MO solution in the bottles functionalized with MLs decayed much faster than that in the bottle without MLs.

Figure \ref{office solar treatment} (d) displays that the enhancement is 9-23 times after 9-12 days. With the extension of light treatment, the dye was also degraded in the bare vial, so from then on the enhancement dropped a little to 7-20 times. Larger MLs on MLs vial\_2 and 3 were more effective, compared to smaller MLs on MLs vial\_1. The results clearly demonstrate that MLs on the inner wall of the bottles could speed up the photodegradation of the dye in water. As the indoor light intensity lacks the UV region (Figure \ref{office solar treatment} (e)), the significant enhancement in photodegradation suggests that bottles functionalized by MLs may be potentially used for light treatment in situations where the sun elevation is low, the weather is cloudy, or the light source is a reflection from snow or local pollution.

\section{Conclusions}
In summary, our work shows the enhancement of photodegradation efficiency with surface MLs occurs under various solution conditions via the same photodegradation pathway. The morphology, number density, and spatial arrangement of MLs have a significant impact on photodegradation efficiency. ML arrays are 100\% more effective than random MLs with the same intensity at focal points and are $\sim 700\% $ more effective than bare surface without MLs. The simulation results suggest that highly ordered ML arrays may result in a locally high concentration of active species around the focal points array, and further accelerate the photodegradation. The photodegradation efficiencies of the dye and three micropollutants in bottles functionalized with MLs were all significantly higher than that in normal bottles, demonstrating the potential application of MLs in the photodegradation of harmful compounds in water. In future work, MLs may be fabricated in commercial drinking water bottles that can be recycled for the decontamination of water.

\section{Acknowledgement}
The authors acknowledge the support from Canada First Research Excellence Fund as part of the University of Alberta’s Future Energy System research initiative and the technical assistance from the Institute for Oil Sands Innovation (IOSI) at the University of Alberta. This research was undertaken, in part, thanks to funding from the Canada Research Chairs Program. This work was also supported by a Natural Sciences and Engineering Research Council of Canada (NSERC) Senior Industrial Research Chair (IRC) in Oil Sands Tailings Water Treatment through the support of Canada’s Oil Sands Innovation Alliance (COSIA), Syncrude Canada Ltd., Suncor Energy Inc., Canadian Natural Resources Ltd., Imperial Oil Resources, Teck Resources Limited, EPCOR Water Services, Alberta Innovates, and Alberta Environment and Parks. MGED, X.W. and Q.X. appreciate funding from NSERC Discovery Grant programs. Q.X. acknowledges Killam Trust-Izaak Walton Killam Memorial Scholarship.
\clearpage

\begin{suppinfo}
Supplementary data to this article can be found in the document attached.
\begin{itemize}
  \item Filename: Supplementary information.docx
\end{itemize}
\end{suppinfo}

\bibliography{Manuscript.bib}
\end{singlespace}
\end{document}